\RequirePackage{fix-cm}
\documentclass[smallextended]{svjour3}       % onecolumn (second format)
\smartqed  % flush right qed marks, e.g. at end of proof
\usepackage{graphicx}
\usepackage{amssymb}
\usepackage{amsmath,amstext,times}
\begin{document}

\title{A Novel Approach for Learning How to Automatically Match Job Offers and Candidate Profiles}

\titlerunning{Matching of Job Offers and Candidate Profiles}        % if too long for running head

\author{Jorge Martinez-Gil \and Alejandra Lorena Paoletti \and Mario Pichler}


\institute{All authors are with\\
              Software Competence Center Hagenberg GmbH\\
              Softwarepark 21\\
              4232 Hagenberg, Austria\\
              %\email{fauthor@example.com}           %  \\
}

\date{Received: date / Accepted: date}

\maketitle

\begin{abstract}
Automatic matching of job offers and job candidates is a major problem for a number of organizations and job applicants that if it were successfully addressed could have a positive impact in many countries around the world. In this context, it is widely accepted that semi-automatic matching algorithms between job and candidate profiles would provide a vital technology for making the recruitment processes faster, more accurate and transparent. In this work, we present our research towards achieving a realistic matching approach for satisfactorily addressing this challenge. This novel approach relies on a matching learning solution aiming to learn from past solved cases in order to accurately predict the results in new situations. An empirical study shows us that our approach is able to beat solutions with no learning capabilities by a wide margin.  
\keywords{Human Resources Management Systems \and Knowledge Engineering \and e-Recruitment}
\end{abstract}

\section{Introduction}
One of the most challenging problems in the Human Resources (HR) domain is to deal with scenarios with a high amount of job applications. This problem has a number of direct and indirect consequences, including but not limited to the waste of resources on processing all these applications. For this reason, researchers and practitioners has focused on finding ways to reduce the cost associated to situations of this kind \cite{key-Mirizzi}\cite{key-Yi}. Additionally, the field of Human Resources carries an old problem on not giving a fair treatment to the job candidates who have spent their time on preparing an application, and however, no feedback on the reasons for not being finally hired for a given position is provided \cite{key-Suerderm}. Organizations do not usually send this feedback to the candidates since this task has little profit for them \cite{key-jikm}. However, we think that if we were able to provide some automatic mechanisms to do so, both sides could benefit, i.e. recruiters can improve their branding reputation, and the unsuccessful applicants can easily know the reasons behind the decisions taken by the recruiters, and take decisions leading to succeed in the future.

To overcome this situation, researchers and practitioners from this field often remark that accurate methods for the automatic matching of applicant profiles and job offers could partly alleviate the problem \cite{key-Malherbe}. Therefore, the design of new automatic approaches that can improve the recruitment processes is an important challenge \cite{key-Malinowski}. Additionally, such an automatic approach could be of great interest for employment agencies and many educational organizations around the world which could try to perform an automatic analysis in order to determine the necessary training courses that could improve the skills and competences of potential unemployed people with respect to the specific needs of a given population segment \cite{key-dcsa}. 

The problem is that current approaches for matching of profiles and job offers are based on a kind of syntactic matching\cite{key-cosrev}. This means that, given a  job offer, existing methods will count the number of requirements that are overlapped within the candidate profiles \cite{key-Mochol}. The major drawback of this approach is that it does not consider the meaning behind every text expression labeling the requirements \cite{key-Paoletti}. Improving this approach requires taking some kind of background or expert knowledge into account. In this context, there are some taxonomies that represent a great source of background knowledge for us. These taxonomies are DISCO\footnote{http://www.disco-tools.eu}, ISCO\footnote{http://www.ilo.org/public/english/bureau/stat/isco/isco08/index.htm} and ISCED\footnote{http://www.uis.unesco.org/Education/Pages/international-standard-classification-of-education.aspx}. These existing taxonomies are of core importance in our work, since we aim to automatically learn how to exploit them for proposing automatic methods which best fits the real hiring needs of the organizations. Therefore, our major contributions are summarized as follows: 

\begin{itemize}
	\item We propose a novel method for the automatic matching between job offers and suitable candidate profiles. This approach relies on a new method for matching learning, and aims for appropriately addressing the traditional problems in this field. 
	\item We test our approach using data from real recruitment scenarios, and we compare our results with OKAPI BM25 algorithm, which represents the state-of-the-art for software solutions based on traditional techniques for information retrieval, i.e. solutions not implementing learning capabilities.
\end{itemize}

The remainder of this work is structured in the following way: Section 2 explains the state-of-the-art on automatic matching between job offers and candidate profiles. Section 3 presents the specific problem we are addressing in this research work. Section 4 explains our proposal for the automatic matching between job offers and candidate profiles. Section 5 shows the result we have achieved by testing our approach in real data from recruitment scenarios. Finally, we draw the conclusions derived from this work. 

\section{State-of-the-art}
The automatic matching of job offers and applicant profiles has already been addressed in the literature \cite{key-Faliagka}\cite{key-Faliagka2}\cite{key-Farber}\cite{key-Kessler}. The question is on which grounds a more realistic technology could be built \cite{key-Tinelli}\cite{key-Tinelli2}. As indicated above some approaches to the problem are grounded in methods from Information Retrieval, i.e. in a nutshell, keyword search methods are conducted on documents representing job offers or curricula vitae, respectively, with keywords characterizing the job seeker or open position, respectively. An unquestionable advantage of methods based on keyword search is that there is no need to bring the raw data, i.e. the texts of job offers or curricula vitae, first into a structured form to be able to apply the matching technology. On the other hand, the methods require manually entry of similarity between key notions into the underlying feature space, which is error-prone and hard to maintain in view of changes and extensions. However, the job market is a very flexible field, where job titles, skill and education concepts, and general terminology are subject to permanent change \cite{key-Paoletti2}.

Alternative approaches favor methods grounded in knowledge bases capturing the terminology used for recruitment \cite{key-Cali}, so that all job and candidate descriptions could be represented by assertional profiles in an adequate knowledge base. As indicated above, matching could be based on likeliness measures defined on filters in knowledge bases. The advantage of these approaches is the flexibility in the matching relations and the perspective that automatic concept classification is supported by knowledge base technology, i.e., the maintenance of terminology will be greatly eased. On the other hand, however, an ontology-based approach requires the definition of a more expressive ontology, i.e. both a language that could be used to define the terminological concepts and their dependencies as well as the assertional profiles, and efficient classification algorithms \cite{key-Garcia}. No satisfactory set-up of a knowledge base for job recruitment exists \cite{key-Racz}.

It is also worth mentioning that there are sector-specific approaches, which can be taken advantage of in order to evaluate skills on either job specifications or candidate profiles. For example in the IT sector, the websites Stack Overflow\footnote{http://www.stackoverflow.com} and GitHub\footnote{http://www.github.com} contain an enormous amount of information about different development skills of a wide range of programmers. There are also domain-specific approaches for trying to match candidates to job-specification and vice-versa \cite{key-Hauff}.

Existing frameworks such as DISCO, ESCO, and ICED, that are meant to capture skill concepts do not fully exploit the opportunities offered by ontologies and their underlying description logics \cite{key-Colucci}. Currently, these frameworks only support taxonomies, i.e. they merely exploit concept subsumption in ontologies, whereas roles (aka attributes) that are used to fine-tune the description are not supported. The first challenge is to fully exploit the capabilities offered by description logics, this is the formal basis of the widely used ontology languages, to capture the terminology used in recruitment applications and to perform reasoning leading to automatically infer some useful facts, e.g. to associate the years of experience with a particular skill such as programming with a particular programming language or assessing marketing studies with a particular software tool.

In general, the methods for matching job offers and candidate profiles are used for addressing one of these research questions:

\begin{itemize}
	\item the ranking of candidates according to their suitability for a certain job offer,
	\item the ranking of job offers according to their suitability for a particular job seeker,
	\item the identification of gaps in candidate profiles hampering successful mediation,
	\item the identification of adequate education and training measures to enhance the employability of job seekers.
\end{itemize}

In this work, we focus on the ranking of candidates according to their suitability for a certain job offer. We consider here the Okapi BM25 \cite{key-lv} as the baseline method to compare a novel approach facing this problem. One of the reasons is that Okapi BM25 could be considered the most accurate computational method using a bag-of-words paradigm \cite{key-lv}. Algorithms of this kind are very popular among existing solutions since working under the bag-of-words paradigm does not require any training phase, and results are usually reasonable. The reason is that documents are seen just as set of words by the software. Therefore, any new proposal should prove its accuracy when compared to it. In this work, we show that an approach using learning capabilities can improve the results from those methods relying in just counting overlapping text expressions.

\section{Definition of the problem}
The research question that we wish to face could be formally described as given a pair of entities $(jo, ap_{i})$ with a score of their likeness $y_{i} \in \mathbb{R}$. The goal is to find a function $f(jo, ap_{i}) \approx  y_{i}$ that approximates for each new entity. For example, $(jo, ap_{i}, y_{i})$, where $jo$ is a job offer, $ap_{i}$ is a list of applicant profiles, and $y_{i}$ is the associated list of scores of each $ap_{i}$ for the job offer $jo$ \cite{key-dcsa}.

In fact, we want to know how can we score the suitability of a number of candidates for a given job offer, assuming the following facts:

\begin{enumerate}
	\item \textit{the ordering of elements belonging to both profiles is not important}, since listing skills, competences, languages, etc. following different orders has no consequences
	\item \textit{the size of each set can be different}, since it is possible for a candidate to have more (or less) skills that those requested by the offer
	\item \textit{some elements from $jo$ can be replaced by some elements from $ap_{i}$ for a certain cost}, since learning C++ after having some expertise of C is not extremely difficult
	\item \textit{for elements cannot be replaced, we have to assume an insertion and deletion cost}, since certain skills, competences or attitudes could be difficult to learn (or to forget)
	\item \textit{some elements could be more important than others}, since the employer could have different levels of priority
	\item \textit{sets can be probably segmented in disjoint subsets of different relevance}: education, skills, languages, social aspects, etc. 
\end{enumerate}

\section{Contribution}
In our opinion, the challenge to be faced has no single solution. The major reason is that every expert evaluating different cases might suggest different ways to face the problem, and therefore, different results will be achieved. Therefore, we propose a machine learning-based adaptive approach. This approach has to be able to compute the cost of transforming a profile into a job offer. Therefore, our approach has to be capable of reproducing results of the human experts when facing this problem. Therefore, for each user or organization that wants to use such a solution, a model must first be trained to capture the know-how or the preferences through a first training stage. Such as model is far from trivial. However, we can envision that a first solution to this problem has to be modeled by making use of the following elements:

\begin{enumerate}
	\item A distance between sets of items
	\item Some background knowledge sources to help determining the replacement costs
	\item The cost of insertion of a new element, and the cost of deletion of unwanted elements, i.e. degree of over-qualification
	\item The way to change the relative importance of the different elements (i.e. multiplicative or an additive way)
\end{enumerate}

It is important to mention that we are going to work with different independent sets (education, competencies, languages, etc.), and therefore, this means that the transformation costs could be different for each independent set. For example, it is not the same to learn a new programming language that acquiring a new official degree. This means that the final costs are calculated by aggregating the partial costs for each independent set.

\subsection{Base distance between sets}
We define the distance between two given sets as the minimum number of basic operations that are necessary to transform an applicant profile into job offer. These basic operations are three: insertions, deletions or substitutions. This distance is useful to compute the cost of transforming a profile into a job offer. More formally, it is possible to describe our distance using the following equation:

\begin{align}
d_{A,B}(i,j) = 
\begin{cases}
  \max(i,j) & \text{ if} \min(i,j)=0, \\
  \min \begin{cases}
          \operatorname{d}_{A,B}(i-1,j) + 1 \\
          \operatorname{d}_{A,B}(i,j-1) + 1 \\
          \operatorname{d}_{A,B}(i-1,j-1) + 1_{(a_i \neq b_j)}
       \end{cases} & \text{ otherwise}
\end{cases}
\end{align}

Let us see an example on how this distance works. Let us assume we have two skill sets representing a job offer and an applicant profile respectively. Our distance is able to compute the minimum amount of basic operations to transform the applicant profile into the job offer. In Fig. 1 we can see an example.

\begin{figure}[h]
	\centering
		\includegraphics[width=0.95\textwidth]{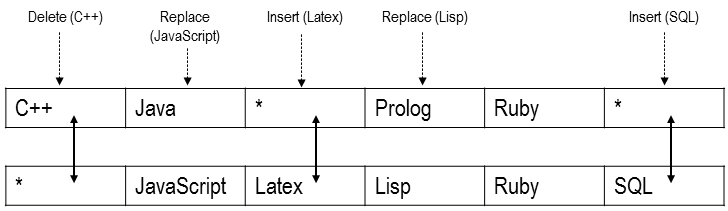}
	\caption{Example of how our approach calculates the distance between sets representing a job offer and applicant profile respectively. We need 5 basic operations from transforming a set into the other one (2 insertions, 2 substitutions, and 1 deletion). The cost associated to each of these basic operations will be determined in the training stage.}
	\label{fig:fi}
\end{figure}

The interesting issue here is that no every human expert gives the same importance to each basic operation. For instance, insertions implies the acquisition of a new skill or competence, deletion implies some kind of miss-qualification, and replacement involves some kind of adaption of the applicant to its new role. This means that by means of analyzing past solved cases, our approach should be able to capture how the expert reason on scenarios on this kind. We will see this in much more detailed way in the next subsections.

\subsection{Replacement costs}
The creation of appropriate knowledge bases covering the relevant terminology from the different thematic areas in recruitment scenarios is a success factor in this approach. The problem is that knowledge bases of this kind do not exist. However, our approach is based on recruiting taxonomies such as ISCO, DISCO, ISCED. These taxonomies are some kind of thesaurus and structured vocabularies for describing skills in a wide range of scenarios such as education, labor market and training respectively. 

These taxonomies give us a comprehensive classification in existing international standards, and represent a terminological basis for the standardized description of skills, competences, professions, vacancies, job requirements, etc., or describe professional degrees, courses of study, and so on \cite{key-dcsa}. 

To illustrate why these taxonomies are important, let us assume that a company is looking for a person who is expert in Java, and they receive the application of a candidate who has a certain degree of mastery of C++. Our approach can calculate the shortest path between Java and C++ in one of these recruitment taxonomies. The replacement cost can be based on the length of this path. In this way, short paths mean low replacement costs and, longer paths mean higher replacement costs. If the path between is too long (according a customizable parameter), then we can compute the deletion and insertion costs.

\subsection{Insertion and deletion costs}
The suitability of a candidate profile for a particular job offer should also consider the minimal cost of insertions and deletions that could transform the profile into the given job offer. These costs must be considered when an applicant profile has not the same number of items than those requested in the job offer or it is not possible to calculate the replacement cost between two different items. 

The calculation of these costs is crucial because it allows us modeling the behavior of the experts who have participated in the first stage of the training. In this context, the insertion costs are an estimate of how much effort the acquisition of a requested item by the potential candidate could cost.

Deletion costs are an estimate of the impact of an item that is not necessary. For instance, an expert might think that candidates with unwanted items could be unmotivated, could request a higher salary or be willing to leave the company as soon as possible. The penalty can be high: if the expert who participates in the training phase tends to punish this factor, neutral; if the expert does not care too much about unnecessary items or it could be even negative; if the expert who trained the model believed that additional items are very useful \cite{key-dcsa}.

In general, it is not a good idea to have a single value for the costs of insertions and deletions in every case. For example, it is usually more difficult to acquire a new degree from an official university than to achieve a certain level of knowledge in a computer language \cite{key-dcsa}. Finally, it is important to remark that the values ​​for these costs are obtained in the training phase since these values ​​are best suited to the specific preferences of the human expert who first trains the model through past solved cases.

\subsection{Weighting approach}
The weighting method defines how to specify the relative importance of the different items within a given job offer. This method allows employers specifying more weight to some factors such as mastery level, years of experience, or simply state relative priorities within a given job offer. This approach could be a be multiplicative approach (an element is twice as important as other elements) or additive (an element is some units far from the other elements). 

\subsection{Overview of the general solution}
After all the partial costs (replacement costs, insertion and deletion costs, and weighting schema) for each of the different categories have been collected, we have to combine all of them following an aggregation strategy. Our aggregation strategy consists of calculating the transformation cost, i.e. given an applicant profile $ap$ and a job offer $jo$, compute the cost that it is necessary to transform $jo$ into $ap$. The formula for computing the final transformation cost (assuming a segmentation in n categories) is the following:

	\[ TC_{d}(jo, ap) \ = \sum^{n}_{i=0} rc^{m}_{i} + ic^{m}_{i} + dc^{m}_{i} \]
	
\begin{center}
\textit{	where d denote the distance defined in (4.1)} \\
\textit{\textit{rc} are the replacement costs, \textit{ic} are the insertion costs, \textit{dc} are the deletion cost, and \textit{m} represents the weighting scheme}
	\end{center}
	
Our goal here is trying to determine the value of each parameter so that the inherent transformation cost that the expert used for solving past cases can be replicated. 

 \[ \exists S_{d,m} = \{ \vec{rc}, \vec{ic}, \vec{dc}\ \} \rightarrow TC_{S} (jo, ap) \approx TC_{H} (jo, ap) \]

	\begin{center}
	\textit{$TC_{S}$ are the costs automatic calculated by our solution, and \\
	$TC_{S}$ are the costs calculated by a human} \\
	\end{center}

In this way, we can automatically elaborate a ranking exactly in the same way the human expert could rank. The approach to follow is simple: the higher the overall transformation cost the worse the position of the candidate in the ranking and vice versa, the lower the transformation cost the better the position of the candidate in the final ranking. 

 \[ Rank: \ \mathbb P \ (\mathbb N \ \times \ ap) \ \rightarrow \ \mathbb P \ (\mathbb N  \ \times \ ap), \ where \ Rank(TC_{i}) \ = \ \vec{TC_{j}} \]

	\begin{center}
	\textit{where $\vec{TC_{j}}$ is given by the order of $\mathbb N$ } \\
	\end{center}

Figure 2 shows an overview of our solution. An initial solution $S_{d,m} = \{\vec{rc}, \vec{ic}, \vec{dc}\}$ is aggregated and produces a ranking of candidates. This ranking is based on the transformation cost from each profile into the given job offer. This ranking has to be compared with the ideal one in order to guide the re-computation processes to a convergence situation. The process is iterative what means that it will be finished either a perfect correlation between an ideal ranking and our ranking is achieved (correlation of 1) or a maximum number of iterations is reached. If we are not able to get a perfect model, then HR experts should decide if the achieved degree of correlation is reasonable for them or a new training stage under different conditions would be needed. 

\begin{figure}[h]
	\centering
		\includegraphics[width=0.7\textwidth]{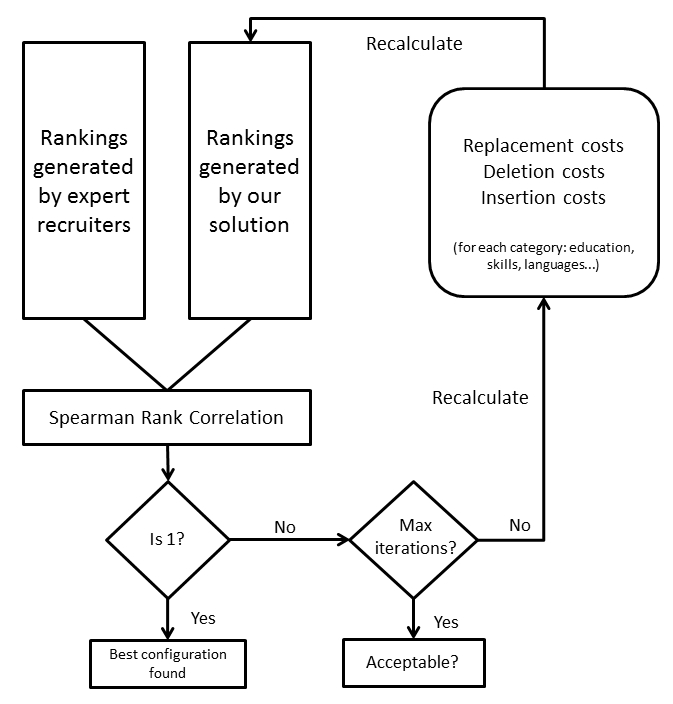}
	\caption{General overview of the proposed solution: our approach looks for a $S_{d,m} = \{\vec{rc}, \vec{ic}, \vec{dc}\}$, so that the ranking generated using $TC_{d}(jo, ap) \ = \sum^{n}_{i=0} rc^{m}_{i} + ic^{m}_{i} + dc^{m}_{i}$ can replicate the ranking made by humans $Spearman \ Rank \ Correlation (R_{TCs}, R_{TCh}) \approx 1$. In this way, $S$ can be reused for automatically solving similar cases in the future}
	\label{fig:learning}
\end{figure}

The rationale behind this approach is that an initial solution should evolve towards the goal of optimizing an aggregation of all the partial costs. This can be done by defining the objective function. In our case, this function is an Spearman rank correlation \cite{key-Deza}, thus a nonparametric measure of statistical dependence between the ideal ranking and the ranking generated by a solution. Since Spearman correlation shows us how well the relationship between two rankings can be described using a monotonic function, the objective we are looking for is a Spearman correlation of 1 that occurs when each of the solutions to be compared is a perfect monotone function of the other. This means that our solution gets exactly the same results than the human expert. Our solution is also scalable, for each new case solved, we can update the values of our aggregation function. Therefore, we are able to learn a function $TC_{S} (jo, ap) \approx TC_{H} (jo, ap)$ that approximates for every new labeled scenario.

\section{Results}
In this section, we are going to show the empirical results we have obtained when working with real data from recruitment scenarios. Please note, for each applicant profile, we have considered only those data concerning: education, skills and, languages. Personal data, photos, references, and so on are out of the scope of this work, and therefore, they have not been considered in this version of our work, but it could be very interesting to handle them as future work.

We are working with samples we have obtained from four major thematic domains: IT, Legal, Logistics, and Marketing. We focus on three kinds of items: Skills, Education and Languages. Table 1 shows some statistical data about the sample we are working with. The left part of the table shows us the average number of items that employers requests, whereas the right part shows us the average number of items that the potential candidates are offering.

\begin{table}[h!]
\centering
\begin{tabular}{l|ccc|ccc}
%\toprule
\hline
          & Skills Req.& Ed. Req. & Lang. Req. & Skills Off. & Ed. Off. & Lang. Off.\\ 
					\hline
IT        & 6.83 $\pm 1.77$           & 0.67   $\pm 0.27$              & 1.00 $\pm 0.00$                & 7.37 $\pm 4.00$          & 0.67 $\pm 0.22$              & 1.00 $\pm 0.00$            \\
Legal     & 5.50  $\pm 2.30$           & 0.67   $\pm 0.27$              & 2.50    $\pm 1.87$            & 3.75  $\pm 1.48$         & 0.83 $\pm 1.29$              & 2.50    $\pm 1.29$         \\
Logistic  & 4.67  $\pm 0.70$           & 0.17   $\pm 0.17$              & 1.17    $\pm 0.17$            & 3.43   $\pm 1.29$         & 0.17 $\pm 0.15$               & 1.33 $\pm 0.57$             \\
Marketing & 5.33  $\pm 3.10$           & 0.33   $\pm 0.27$             & 1.33      $\pm 0.67$          & 3.75  $\pm 2.50$          & 0.67 $\pm 0.29$               & 1.63  $\pm 1.27$            \\ 
\hline
\end{tabular}
\caption{Summary of the data we are going to work with. The three first numeric columns represent the number of items that employers request. The rest represent the number of items that candidates offer. All figures represent average means and variances}
\label{my-label}
\end{table}

The rationale behind these experiments is to assess if our training model is able to replicate well enough the behavior from human experts when ranking candidates and to beat the baseline approach in this context. The challenge is so that given a number of candidates and number of solved cases from the past (job offers and its associated candidate's rankings), trying to predict the rankings for new job offers so that these automatically generated rankings can perfectly replicate the rankings made by human experts.

\begin{figure}[h]
	\centering
		\includegraphics[width=0.60\textwidth]{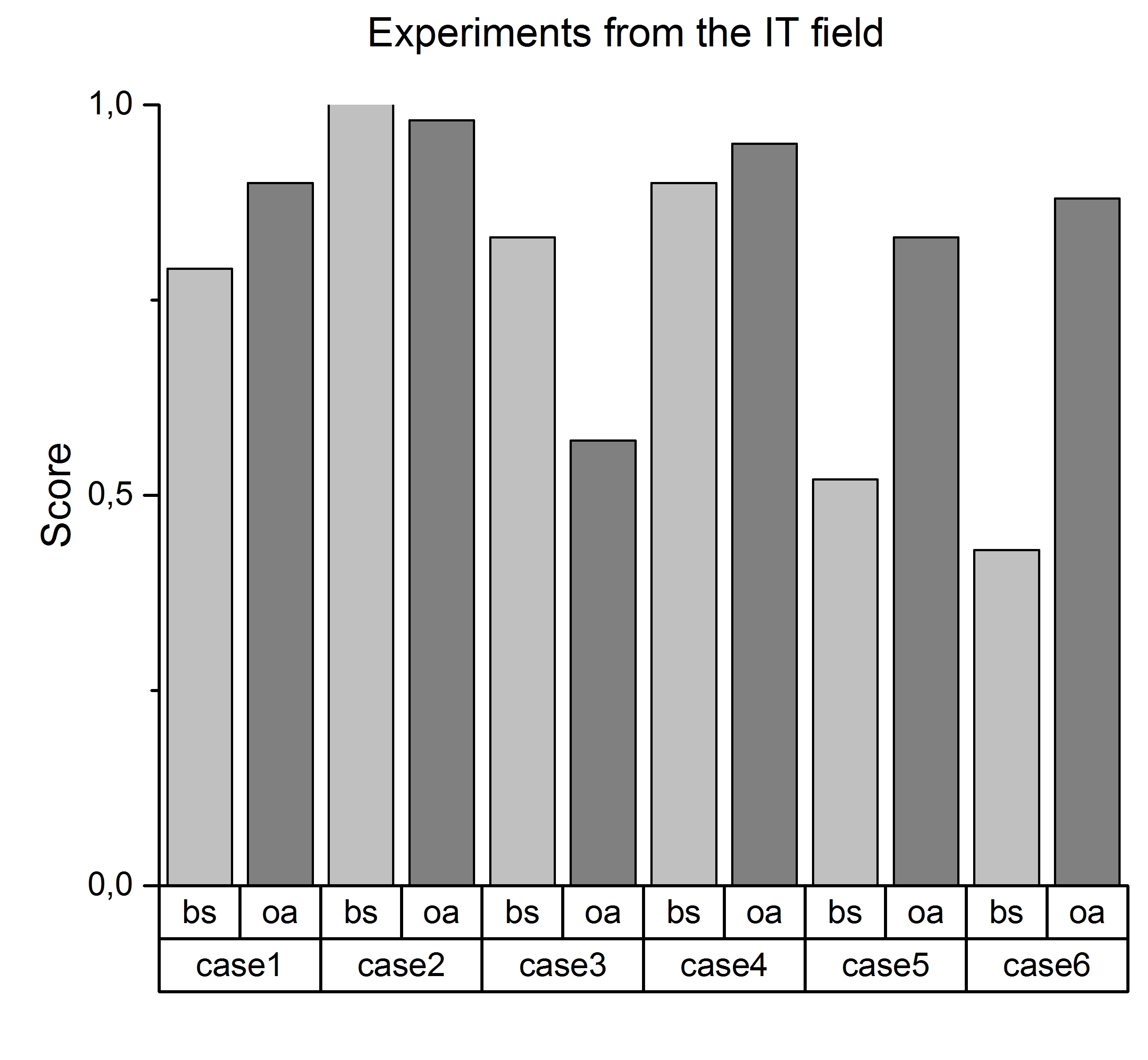}
	\caption{Results obtained in the IT field. Our approach (oa) outperforms the baseline (bs) in 4 out 6 cases}
	\label{fig:it}
\end{figure}

The results we have achieved are represented in Fig. 3 (IT domain), Fig 4. (Legal domain), Fig. 5 (Logistics domain), and Fig. 6 (Marketing domain). After training our model with solved cases from an expert, and given eight candidates for each of the thematic domains, we predict the ranking for a new job offer in the way the human expert would make. 

We got six solved cases for each of the four thematic domains. These results show the degree of correlation between the predictions made by our model and the real ranking made by the expert for each of these new job offers. It is important to remark that the higher amount of candidates to rank the more difficult to be completely accurate, since we get a combinatorial explosion of possible rankings. Finally, in order to get a baseline for the problem, we have executed the algorithm Okapi BM25 \cite{key-lv} over the same scenario to see if our learning mechanism really produces any benefit. \\

\begin{figure}[h]
	\centering
		\includegraphics[width=0.60\textwidth]{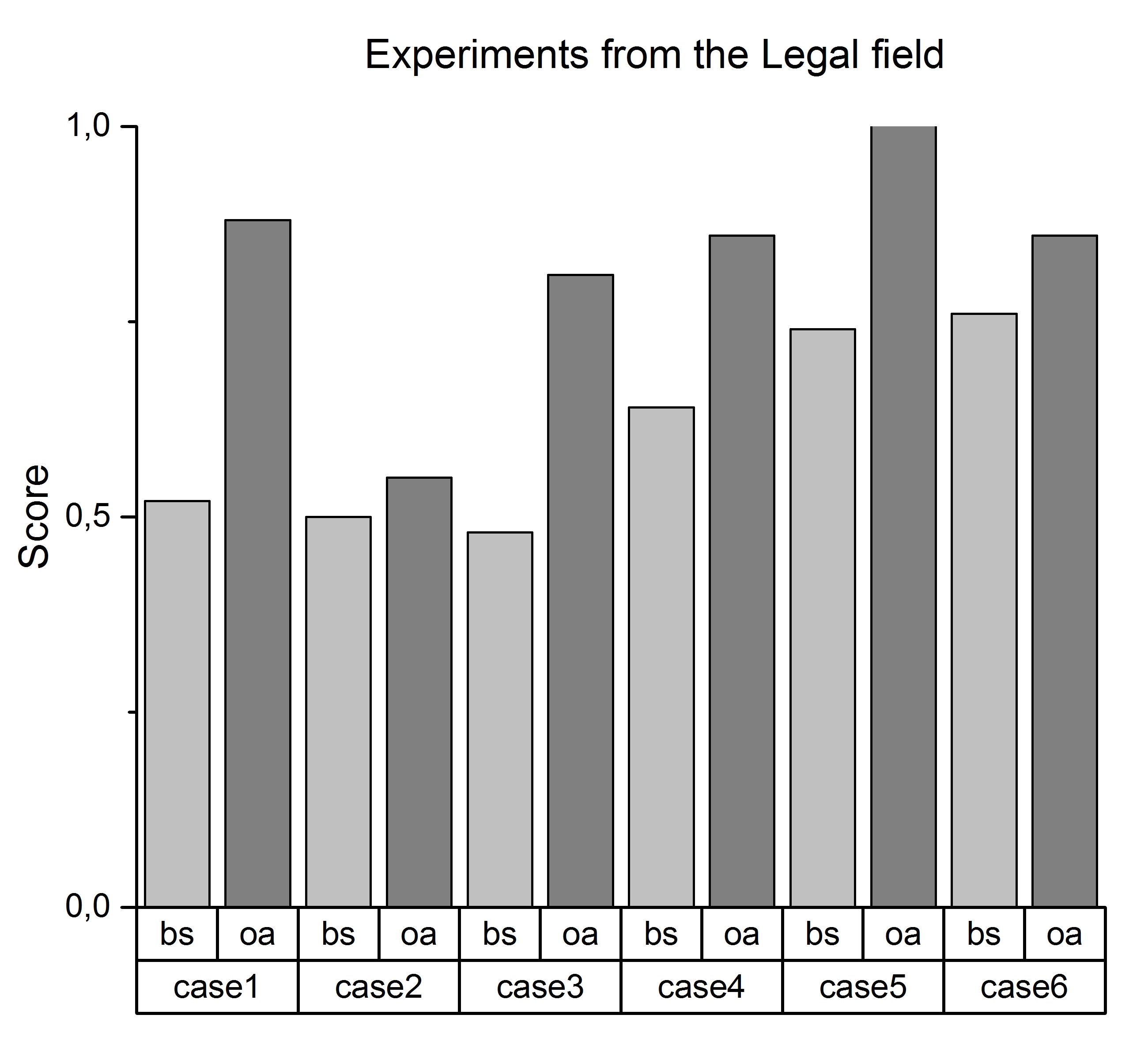}
	\caption{Results obtained in the Legal field. Our approach (oa) outperforms the baseline (bs) in all cases}
	\label{fig:legal}
\end{figure}

Our solution is not able to always beat the baseline since that knowledge from experts is not always consistent. This means that experts are extremely accurate for determining the first positions of a ranking, but once they know some candidates are not suitable for a given job offer, they do not rank them consistently. This case is fatal for our training model since it relies on the total consistency of solved cases.

\begin{figure}[h]
	\centering
		\includegraphics[width=0.60\textwidth]{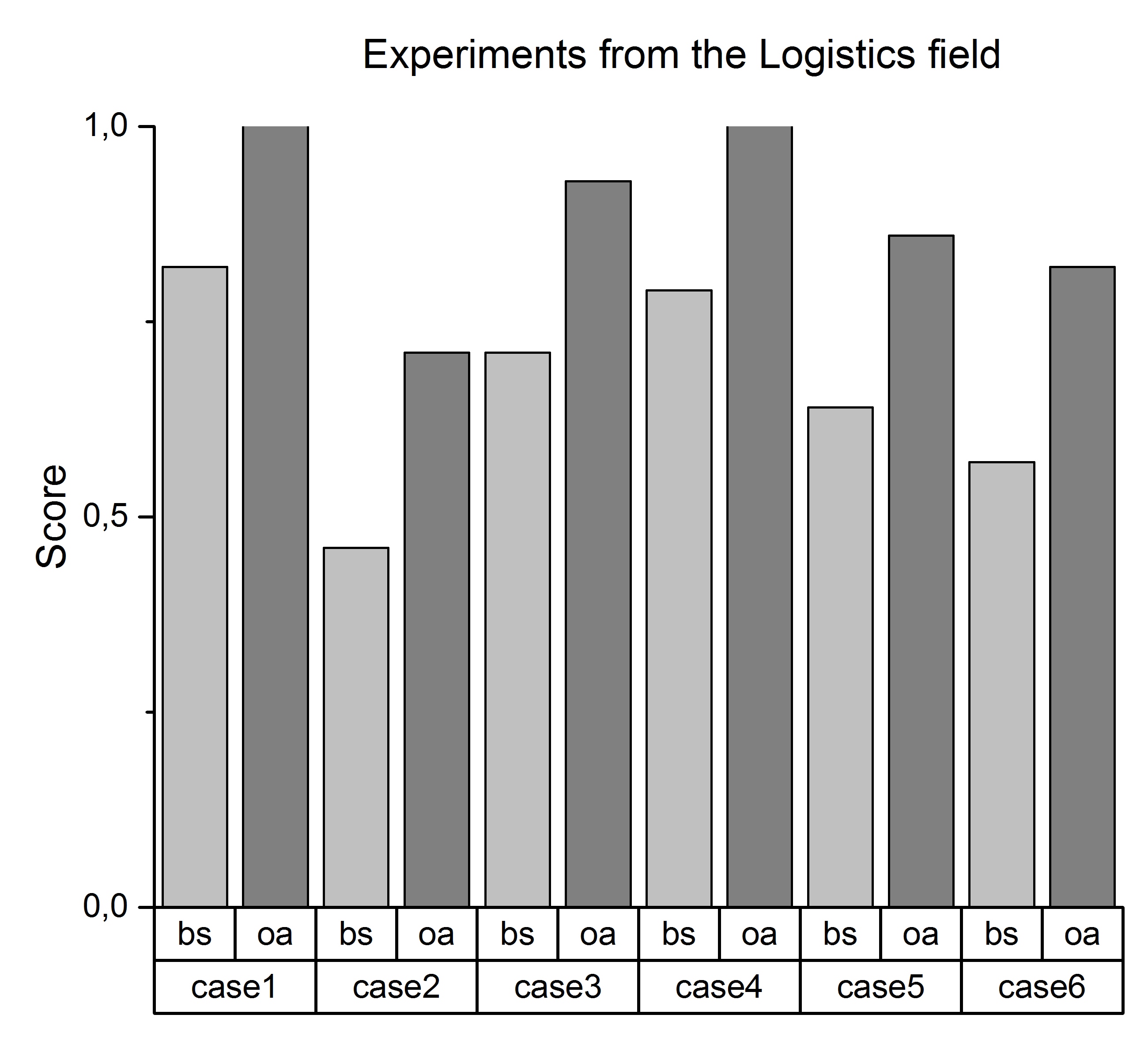}
	\caption{Results obtained in the Logistics field. Our approach (oa) outperforms the baseline (bs) in all cases}
	\label{fig:logistics}
\end{figure}

\begin{figure}[h]
	\centering
		\includegraphics[width=0.60\textwidth]{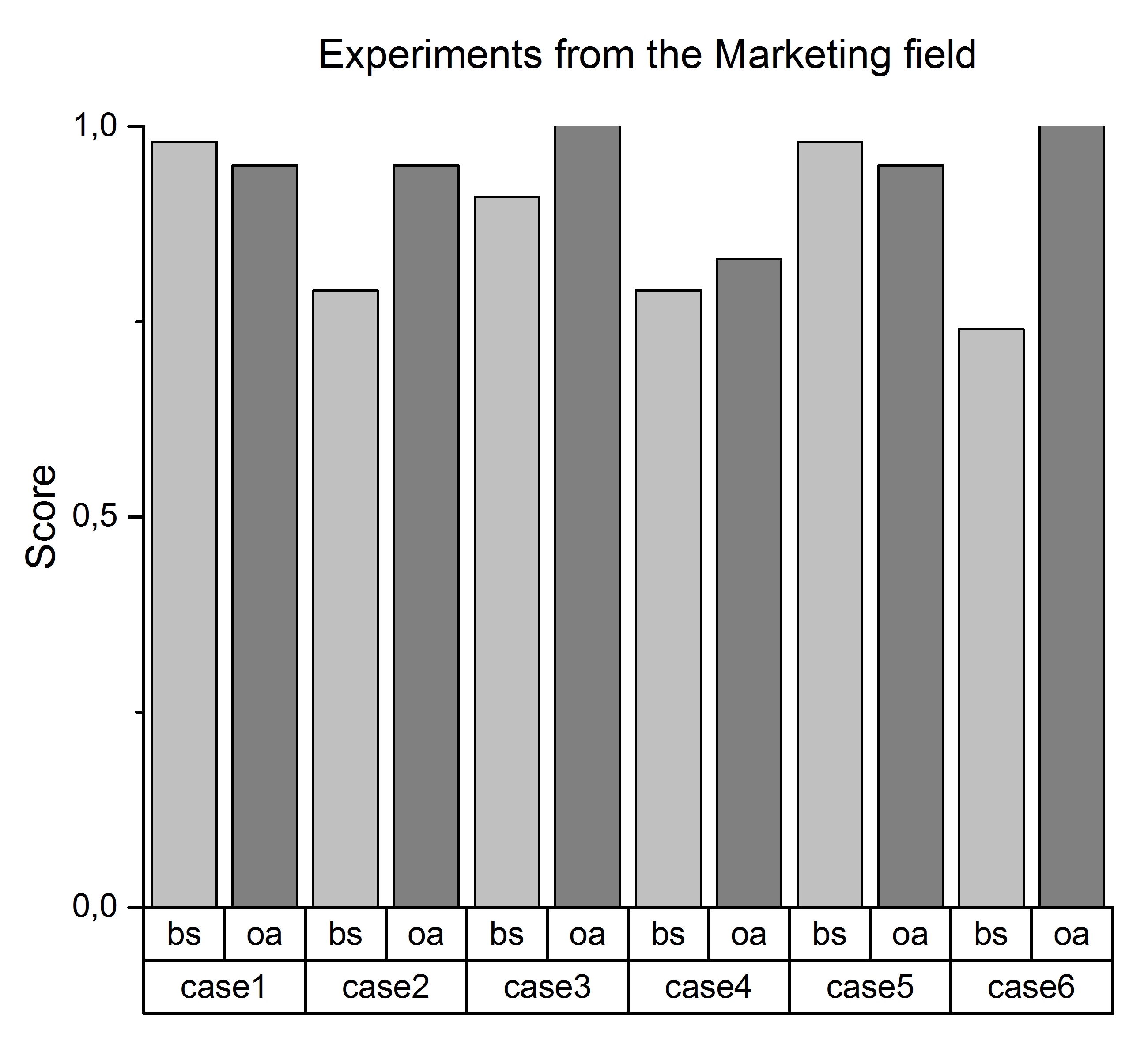}
	\caption{Results obtained in the Marketing field. Our approach (oa) outperforms the baseline (bs) in 4 out 6 cases}
	\label{fig:marketing}
\end{figure}

We have also performed a significance test leading to investigate whether we can accept or reject the null hypothesis on our experiments. We estimated our threshold for the p-value parameter as $5.0 \cdot 10^{-2}$. This means that achieving a statistically significant Spearman rank-order correlation can be interpreted so that there is less than a 5\% chance that the strength of the correlation happened by chance, and therefore, the correlation can be considered as statistically significant. Table 2 shows the associated matrix of statistical significance.

\begin{table}
	\centering
    \begin{tabular}{|l|c|c|c|c|c|c|c|}
		\hline
    Domain         & case1   & case2   & case3 & case4 & case5 & case6 \\
		\hline
    IT        & $1.1 \cdot 10^{-3}$ 	 & $1.00 \cdot 10^{-6}$		& $7.0 \cdot 10^{-2}$*& $1.5 \cdot 10^{-4}$ & $5.3 \cdot 10^{-3}$ & $2.0 \cdot 10^{-3}$ \\
    Legal    & $2.0 \cdot 10^{-3}$   & $7.9 \cdot 10^{-2}*$    & $7.4 \cdot 10^{-3}$  & $3.1 \cdot 10^{-3}$ & ${0.0}$ & $3.1 \cdot 10^{-3}$ \\
    Logistic  & $0.0$   & $2.4 \cdot 10^{-2}$    & $4.0 \cdot 10^{-4}$  & 0.0 & $2.0 \cdot 10^{-3}$ & $6.3 \cdot 10^{-3}$ \\
    Marketing & $1.5 \cdot 10^{-4}$   & $1.5 \cdot 10^{-4}$    & 0.0  & $5.3 \cdot 10^{-3}$ & $1.5 \cdot 10^{-4}$  & 0.0  \\
		\hline
    \end{tabular}
		\caption{Matrix of statistical significance. Values below $5.0 \cdot 10^{-2}$ show a significant correlation. Values above are marked with an asterisk}
\end{table}

Finally, Table 3 shows the time needed for delivering each result. The experiments were performed on a computer using a Intel(R) Core(TM) i7-4790 CPU @ 3.60Ghz over Windows 7 Professional. This time includes the training phase and the elaboration of the ranking. We exclude here a) the load of the knowledge in main memory, since this has to be done just once, and b) the comparison of the given result and the expected one by using the Spearman rank-order correlation method, since we assume this issue might not be offered as functionality in a production environment.

As we can see, the time needed for delivering each result is of the same magnitude for each of the four particular scenarios from the different thematic fields. However, the time needed for delivering each result cannot be just the same for each of the experiments, since some cases need to automatically process larger job offers, or larger applicant profiles, or just trying to compute the replacement costs, that it is by far, the most expensive operation in terms of computational time, since it requires to find a viable path between two skills into the knowledge base.

\begin{table}
	\centering
    \begin{tabular}{|l|c|c|c|c|c|c|c|}
		\hline
    Domain         & case1   & case2   & case3 & case4 & case5 & case6 \\
		\hline
    IT jobs        & 82 & 87 & 76 & 75 & 76 & 80 \\
    Legal jobs     & 71 & 71 & 75 & 79 & 79 & 72 \\
    Logistic jobs  & 63 & 63 & 66 & 63 & 62 & 64 \\
    Marketing jobs & 73 & 73 & 72 & 72 & 71 & 74 \\
		\hline
    \end{tabular}
		\caption{Time needed for delivering each result expressed in seconds}
\end{table}

\subsection{Discussion}
Finally, it is important to note that results that we have achieved show that our approach has beaten the baseline method in 20 out of 24 cases. Therefore, our approach seems to be promising in this context as we have proved to beat the baseline in the experiments performed. In fact, we think our novel approach presents some positive aspects over the existing solutions for the automatic matching of job offers and applicant profiles. Some of these advantages have been already envisaged in \cite{key-jikm}, and can be summarized as follows:

\begin{itemize}
	\item Reduction of the efforts (in terms of cost and time) to find appropriate links between job offers and applicant profiles: This fact is especially positive in organizations with a great volume of hiring activity. 
	\item Improving the traditional matching of offers and profiles: This improvement represents an advantage since it is not only about giving more chances to good candidates, but about helping recruiters to identify talent that may remain hidden too.
	\item Elimination of the need for professionals from the HR sector to have specific knowledge concerning a professional field or skill set: This is due to the fact that our approach can work with past solved cases from a number of professional domains. 
	\item Possibility of providing feedback to non-selected candidates by means of a trace corresponding to the execution of the algorithm over the candidate’s data.
\end{itemize}

\section{Conclusions and Future Work}
In this work, we have introduced a novel method for the accurate and automatic matching of job offers and suitable candidate profiles in an automatic way. The aim of this new method is to reduce the effort (in terms of cost and time) to find the most suitable applicant profiles for a given job offer. This solution is able to learn how human experts solved cases in the past, in order to predict how they would behave in a future situation. The model we have proposed is based on automatically computing transformation costs by using background knowledge (in the form of well-known taxonomies). In this way, it is possible to us to evaluate the suitability of a job applicant for a given job offer in the way an human expert would do it.

Our approach leads to positive results when solving real recruitment cases. In this context, our approach is able to avoid some of the traditional problems in the field of automatic job and profile matching. With respect to uncertainty derived from the use of plain text when describing job offers and applicant profiles, our approach suggests representing profiles and job offers using shared terminologies in order to overcome the problem of dealing with heterogeneous representations of the skill or competence. With regard to the inability of existing approaches to leverage external knowledge, our new method tries to exploit external sources to make estimates on the cost of acquiring a new skill or competence. With regard to the design of a improved method for matching job offers and candidate profiles, our approach proposes a new method that involves collecting a wide range of partial measures, which can be strategically combined to replicate the behavior of the experts. As a result, we have a solution with a positive impact on organizations or users with a high volume of hiring needs.

As a future work, we plan to design a user-friendly solution that can manage the problem of automatically mapping the information contained in applicant profiles to the frameworks DISCO, ISCO and ISCED. The rationale behind such solution is that employers and job seekers can have greater flexibility when creating their job offers and applicant's profiles respectively by choosing among a plethora of different document formats. In this way, the whole recruiting process can be more comfortable for both sides.  

\section*{Acknowledgments}
The authors would like to thank the anonymous reviewers for their helpful and constructive comments that greatly contributed to improve this work. The research reported in this paper was partially supported by the Austrian Forschungsforderungsgesellschaft (FFG) with the Bridge Project “Accurate and Efficient Profile Matching in Knowledge Bases” (ACEPROM) under contract [FFG: 841284]. The research reported in this paper has been partially supported by the Austrian Ministry for Transport, Innovation and Technology, the Federal Ministry of Science, Research and Economy, and the Province of Upper Austria in the frame of the COMET center SCCH [FFG: 844597]. 

%\newpage


\begin{thebibliography}{5}

\bibitem{key-Cali} Cali, A., Calvanese, D., Colucci, S., Di Noia, T., Donini, F.M.: A Logic-Based Approach for Matching User Profiles. Proc. of KES 2004: 187-195.

\bibitem{key-Colucci} Colucci, S., Di Noia, T., Di Sciascio, E., Donini, F.M., Mongiello, M., Mottola, M.: A Formal Approach to Ontology-Based Semantic Match of Skills Descriptions. J. UCS 9(12): 1437-1454 (2003).

\bibitem{key-Deza} Deza, M.M., Deza, E.: Encyclopedia of Distances. Springer (2013).

\bibitem{key-Faliagka} Faliagka, E., Tsakalidis, A.K., Tzimas, G.: An Integrated E-Recruitment System for Automated Personality Mining and Applicant Ranking. Internet Research 22(5): 551-568 (2012).

\bibitem{key-Faliagka2} Faliagka, E., Iliadis, L.S., Karydis, I., Rigou, M., Sioutas, S., Tsakalidis, A.K., Tzimas, G.: On-line consistent ranking on e-recruitment: seeking the truth behind a well-formed CV. Artif. Intell. Rev. 42(3): 515-528 (2014).

\bibitem{key-Farber} Farber, F., Weitzel, T., Keim, T.: An Automated Recommendation Approach to Selection in Personnel Recruitment. Proc. of AMCIS 2003: 302.

\bibitem{key-Garcia} Garcia Sanchez, F., Martinez-Bejar, R., Contreras, L., Fernandez-Breis, J.T., Castellanos Nieves, D.: An ontology-based intelligent system for recruitment. Expert Syst. Appl. 31(2): 248-263 (2006).

\bibitem{key-Hauff} Hauff, C., Gousios, G.: Matching GitHub Developer Profiles to Job Advertisements. MSR 2015: 362-366.

\bibitem{key-Kessler} Kessler, R., Béchet, N., Roche, M., Torres-Moreno, J.M., El-Bèze, M.: A hybrid approach to managing job offers and candidates. Inf. Process. Manage. 48(6): 1124-1135 (2012).

\bibitem{key-Malherbe} Malherbe, E., Cataldi, M., Ballatore, A.: Bringing Order to the Job Market: Efficient Job Offer Categorization in E-Recruitment. Proc. of SIGIR 2015: 1101-1104.

\bibitem{key-Malinowski} Malinowski, J., Keim, T., Wendt, O., Weitzel, T.: Matching People and Jobs: A Bilateral Recommendation Approach. Proc. of HICSS 2006.

\bibitem{key-jikm} Martinez-Gil, J.: An Overview of Knowledge Management Techniques for e-Recruitment. JIKM 13(2) (2014).

\bibitem{key-cosrev} Martinez-Gil, J. Aldana-Montes, J.F.: Semantic similarity measurement using historical google search patterns. Information Systems Frontiers 15(3): 399-410 (2013).

\bibitem{key-dcsa} Martinez-Gil, J., Paoletti, A.L., Schewe, K.D.: A Smart Approach for Matching, Learning and Querying Information from the Human Resources Domain. Proc. ADBIS (Short Papers and Workshops) 2016: 157-167.

\bibitem{key-Mirizzi} Mirizzi, R., Di Noia, T., Di Sciascio, E., Trizio, M.: A Semantic Web Enabled System for Resume Composition and Publication. Proc. of ICSC 2009: 583-588.

\bibitem{key-lv} Lv, U., Zhai, C.: A Log-Logistic Model-Based Interpretation of TF Normalization of BM25. Proc. of ECIR 2012: 244-255.

\bibitem{key-Mochol} Mochol, M., Wache, H., Nixon, L.J.B.: Improving the Accuracy of Job Search with Semantic Techniques. Proc. of BIS 2007: 301-313.

\bibitem{key-Racz} Racz, G., Sali, A., Schewe, K-D.: Semantic Matching Strategies for Job Recruitment: A Comparison of New and Known Approaches. Proc. of FoIKS 2016: 149-168.

\bibitem{key-Paoletti} Paoletti, A.L., Martinez-Gil, J., Schewe, K.D.: Extending Knowledge-Based Profile Matching in the Human Resources Domain. Proc. of DEXA (2) 2015: 21-35.

\bibitem{key-Paoletti2} Paoletti, A.L., Martinez-Gil, J., Schewe, K.D.: Top-k Matching Queries for Filter-Based Profile Matching in Knowledge Bases. Proc. of DEXA (2) 2016: 295-302.

\bibitem{key-Suerderm} Suerdem, A., Oztaysi, B.: Collaborative Requirement Prioritization for an E-Recruitment Platform for Qualified but Disadvantaged Individuals. Proc. of KICSS 2014: 547-556.

\bibitem{key-Tinelli} Tinelli, E., Cascone, A., Ruta, M., Di Noia, T., Di Sciascio, E., Donini, F.M.: I.M.P.A.K.T.: An Innovative Semantic-based Skill Management System Exploiting Standard SQL. Proc. of ICEIS (2) 2009: 224-229.

\bibitem{key-Tinelli2} Tinelli, E., Colucci, S., Donini, F.M., Di Sciascio, E., Giannini, S.: Embedding semantics in human resources management automation via SQL. Appl. Intell. 46(4): 952-982 (2017)

\bibitem{key-Yi} Yi, X., Allan, J., Croft, W.B.: Matching resumes and jobs based on relevance models. SIGIR 2007: 809-810.

\end{thebibliography}
\end{document}